%% file: spin-logic.tex
\documentclass[doublecol]{epl2} 

\usepackage[utf8]{inputenc}
\usepackage{verbatim}
\usepackage{enumerate}
\usepackage{graphicx}
\graphicspath{{pics/}}
\usepackage{hyperref}
\hypersetup{
  pdftitle={Ground State Spin Logic},
  colorlinks,%
  citecolor=black,%
  filecolor=black,%
  linkcolor=black,%
  urlcolor=black,
  pdfauthor={James Daniel Whitfield, Mauro Faccin, and Jacob Biamonte},
  pdfsubject={Optimization, Ising Hamiltonian, adiabatic quantum computing, ground state computing, pseudo-boolean functions, circuits, adder circuits, SR latch},
}
\usepackage{color}
\usepackage{subfigure}
\usepackage{booktabs}

\newcommand{\ket}[1]{|#1\rangle}
\newcommand{\Span}{\textrm{span}}

\newcommand{\bra}[1]{\langle#1|}

\newcommand{\id}{\mathbf{1}}
\newcommand{\LL}{\mathcal{L}}
\def\AND{{\sf AND}}
\def\NAND{{\sf NAND}}
\def\COPY{{\sf COPY}}
\def\NOR{{\sf NOR}}
\def\OR{{\sf OR}}
\def\XOR{{\sf XOR}}

\def\EQUIV{{\sf EQUIV}}

\newcommand{\eqref}[1]{(\ref{#1})}

\renewcommand{\thesubfigure}{\alph{subfigure}}
\makeatletter
  \renewcommand{\@thesubfigure}{\thesubfigure)\space}
  \renewcommand{\p@subfigure}{\thefigure}
\makeatother

\begin{document}

\title{Ground State Spin Logic}

\author{J. D. Whitfield\inst{1,2,3}\thanks{E-mail: \email{jdw2168@columbia.edu}} 
\and M. Faccin\inst{2}\thanks{E-mail: \email{mauro.faccin@isi.it}}
\and J. D. Biamonte \inst{2,4}\thanks{E-mail: \email{jacob.biamonte@qubit.org}}
}
\shortauthor{J. D. Whitfield \etal}

\institute{                    
  \inst{1} Department of Physics, Columbia University, 538 W.~120th St, New York, NY 10027, USA\\
  \inst{2} Institute for Scientific Interchange, Via Alassio 11/c, 10126 Torino, Italy \\
  \inst{3} Quantum Information Technologies, NEC Laboratories America, 4 Independence Way, Princeton, NJ 08540, USA\\
  \inst{4} Centre for Quantum Technologies, National University of Singapore, Block S15, 3 Science Drive 2, Singapore 117543
}


\abstract{
Designing and optimizing cost functions and energy landscapes is a problem encountered in many fields of science and
engineering. These landscapes and cost functions can be embedded and annealed in 
experimentally controllable spin Hamiltonians. 
Using an approach based on group theory and symmetries, we examine the embedding of Boolean logic gates into the 
ground state subspace of such spin systems. We describe parameterized families of diagonal Hamiltonians and symmetry operations which
preserve the ground state subspace encoding the truth tables of Boolean formulas. 
The ground state embeddings of adder circuits are used 
to illustrate how gates are combined and simplified using symmetry.
Our work is relevant for experimental demonstrations of ground state embeddings 
found in both classical optimization as well as adiabatic quantum optimization.
}
\maketitle

The embedding of energy landscapes into the ground 
state subspace of spin systems is a task commonly encountered 
in both classical~\cite{Hartmann05,Bar82,KGV83} and 
quantum optimization~\cite{Ohzeki11,Das2008,Das05}.
Finding a state in this subspace is equivalent to a wide variety 
of NP-complete decision problems and NP-hard optimization problems~\cite{Bar82,Hartmann05,Altshuler09,Altshuler10,Choi11,Dickson11a} which
have received renewed interest in the wake of adiabatic
quantum computation~\cite{Ohzeki11,Apolloni89,Amara93,Finnila94,Kadowaki98,Farhi01} and its experimental realizations~\cite{Britton12,Johnson11, 2010PhRvB..82b4511H}.
Recent works have focused on embedding cost functions into the ground state subspace of spin 
systems~\cite{JDB08,Gu12,Crosson10,Garnerone11,Pudenz11,Perdomo08,Choi11,Das2008,Apolloni89,Rosenbaum,Amara93,Finnila94,Kadowaki98,Farhi01}
and cellular automata~\cite{Lent94,Gu09,Crosson10,Burgarth11}.
While the emphasis and techniques used in previous work varies,
many of the fundamental results overlap. 

In this letter, we use symmetries of Boolean functions to unify and extend 
various constructions of Hamiltonians embedding Boolean functions into their
ground state subspaces.  We perform a systematic analysis of the Hamiltonians
embedding all two-input, one-output gates using our group theoretic approach. 
We also report on a new family of Hamiltonians embedding the universal 
logic gate \NAND\ and present a new \XOR\ Hamiltonian embedding which 
encompass several previous results~\cite{JDB08,Gu12,Rosenbaum}. Both of our
constructions have three free parameters providing previously ignored degrees 
of freedom which could be useful when considering experimental constraints.
Extensions of our symmetry arguments to larger Boolean functions are 
demonstrated using adder circuits of increasing complexity. 

While we focus on embedding circuits into the ground state, 
the application of symmetry arguments is quite 
general and can be used in the construction of Hamiltonians 
for other embedding problems recently studied in adiabatic quantum computing 
such as lattice protein folding~\cite{Perdomo08,Perdomo12}, adiabatic quantum 
simulation~\cite{Biamonte11}, machine learning~\cite{Pudenz11},
or search engine rankings~\cite{Garnerone11}. 

Throughout this letter, we use diagonal
Hamiltonians of $N$ spins 
\begin{equation}
  H = \sum_i c_i \sigma_i+\sum_{ ij} c_{ij}
  \sigma_i\sigma_j+\sum_{ijk} c_{ijk}\sigma_i\sigma_j\sigma_k+...
\label{eq:H1}
\end{equation}
with $\sigma \equiv \sigma^z$ defined by
  $\sigma=\ket{0}\bra{0}-\ket{1}\bra{1}$.
Since the eigenvalues of $\sigma$ are $\pm 1$, we identify Boolean 
variable, $x\in\{0,1\}$, with $(\id-\sigma)/2$ instead of $\sigma$ itself.
The subscript of each $\sigma$ indicates which spin the operator acts on. Terms
such as $\sigma_i\sigma_j$ are understood as the tensor product $\sigma_i\otimes\sigma_j$.

Limiting the Hamiltonian in eq.~\eqref{eq:H1} to 
two-spin interactions yields the 
experimentally relevant~\cite{Britton12,2010PhRvB..82b4511H,Johnson11}
tunable Ising Hamiltonian which will be our primary focus.

The idea of ground state spin logic is to embed Boolean functions, $f:\{0,1\}^n\rightarrow\{0,1\}^m$, into the ground state
subspace, $\LL(H_{f(\mathbf{x})})$, of spin Hamiltonian 
$H_{f(\mathbf{x})}(\sigma_i,\sigma_j,\cdots,\sigma_k)$
acting on the spins $\sigma_i$, $\sigma_j$, \ldots, $\sigma_k$.
As an example, consider the universal \NAND~gate defined by 
$\NAND(x,y)=\bar x\lor\bar y$.
The corresponding Hamiltonian, $H_{\bar x\lor\bar y}(\sigma_1,\sigma_2,\sigma_3)$, should
have the following ground state subspace
\begin{eqnarray}
	\LL(H_{\bar x\lor\bar y})
&=& \Span\{\ket{x}\ket{y}\ket{{\bar{x}\lor \bar{y}}}\}\\
&=& \Span\{\ket{001},\ket{011},\ket{101},\ket{110}\}\nonumber
\end{eqnarray}
Using the $\sigma$ matrices, such a Hamiltonian is given in \cite{Crosson10} as, 
\begin{equation}
	H_{\bar{x}\lor\bar{y}}(\sigma_1,\sigma_2,\sigma_3)= 2\id+\left( \id + \sigma_1+\sigma_2 -\sigma_1\sigma_2\right)\sigma_3
\label{eq:nand1}
\end{equation}
This construction uses a three-spin interaction which can be
replaced using the same number of spins and only two-spin 
interactions. This was done in~\cite{JDB08,Gu12} by penalizing 
and rewarding certain interactions such that the ground state subspace is not 
altered while the higher energy eigenstates are.  

Now we introduce the first result of our paper: a three-parameter family of 
Hamiltonians that generalizes the formulas found in~\cite{JDB08,Gu12,Rosenbaum}
and elsewhere.
Using coefficients labeled as in eq.~\eqref{eq:H1}, 
the constraint that one eigen-subspace is four-fold degenerate 
and contains states $\ket{001}$, $\ket{011}$, $\ket{101}$, 
and $\ket{110}$ leads to the following three equalities:
\begin{eqnarray}
	c_3&=& c_1+c_2\\
	c_{13}&=& c_{12}+c_1\\
	c_{23}&=& c_{12}+c_2
\end{eqnarray}
After enforcing these constraints, the energies are
\begin{eqnarray}
	E_{degen}&=& -c_1-c_2-c_{12}\\
	E_{000}&=& 3(c_1+c_2+c_{12})\\
	E_{010}&=& 3c_1-c_2-c_{12}\\
	E_{100}&=& 3c_2-c_1-c_{12}\\
	E_{111}&=& 3c_{12}-c_1-c_2
\end{eqnarray}
For $c_1$, $c_2$, and $c_{12}$ greater than zero, the degenerate
space is always the ground state.  In closed form the three-parameter
family of Hamiltonians encoding \NAND\ in the ground state is
\begin{eqnarray}
	H_{\bar{x}\lor\bar{y}}(\sigma_1,\sigma_2,\sigma_3)&=&(c_1\sigma_1+c_2\sigma_2)(\id+\sigma_3)\label{eq:hnand}\\
			&&+(c_1+c_2)\sigma_3+c_{12}\sum_{i<j}\sigma_i\sigma_j\nonumber
\end{eqnarray}
with $c_1,c_2,c_{12}>0$.  The freedom to select these parameters could be 
desirable as it reduces the constraints placed on an experimental realization.

The ground state energy of the \NAND{} Hamiltonian, is $-(c_1+c_2+c_{12})$ 
instead of zero. Some authors choose to consider positive 
semi-definite Hamiltonians, however the addition of multiples of the identity 
does not alter energy differences within the landscape of the problem and we
choose not to enforce this constraint.

As the \NAND\ gate is universal for the construction of
logic circuits, the NP-complete problem CIRCUIT-SAT, where the question ``Is there 
an input corresponding to the output of logical one?'' is embedded using only positive
couplings and positive local fields.  This leads to an alternative proof that finding 
the ground state of spin Hamiltonians with anti-ferromagnetic couplings in a 
magnetic field is NP-hard~\cite{Bar82}.

\begin{figure}[tb]
\centering
\subfigure[]{
\includegraphics[width=0.45\columnwidth]{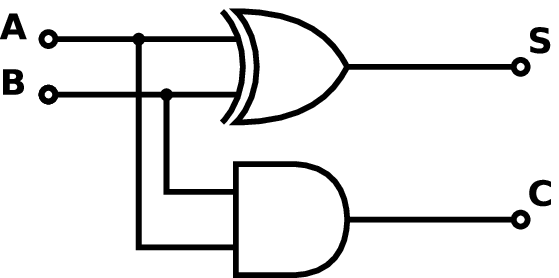}
\label{fig:halfadder-circuit}}
\subfigure[]{
\includegraphics[width=0.45\columnwidth]{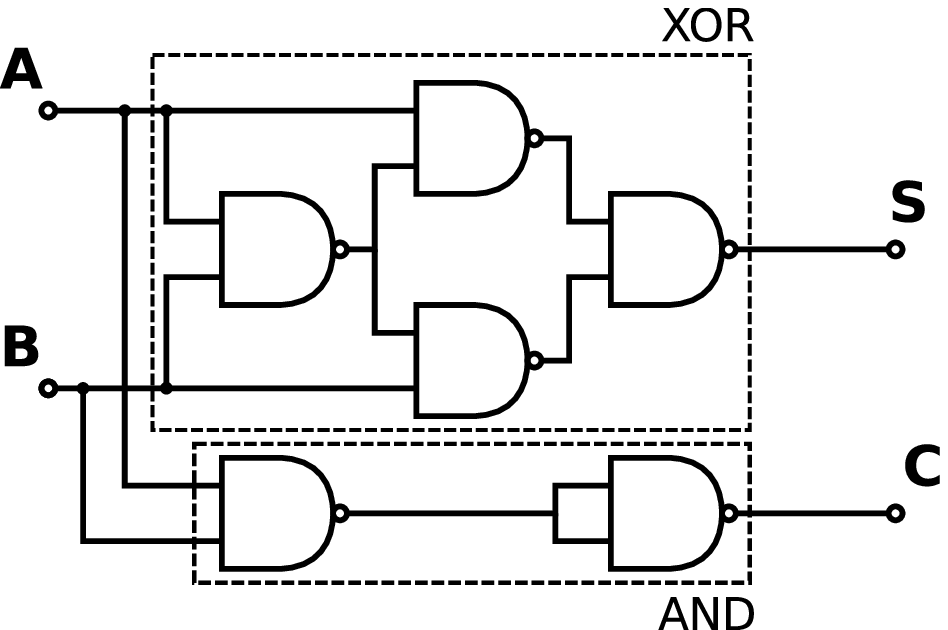}}\\
\subfigure[]{
\includegraphics[width=0.45\columnwidth]{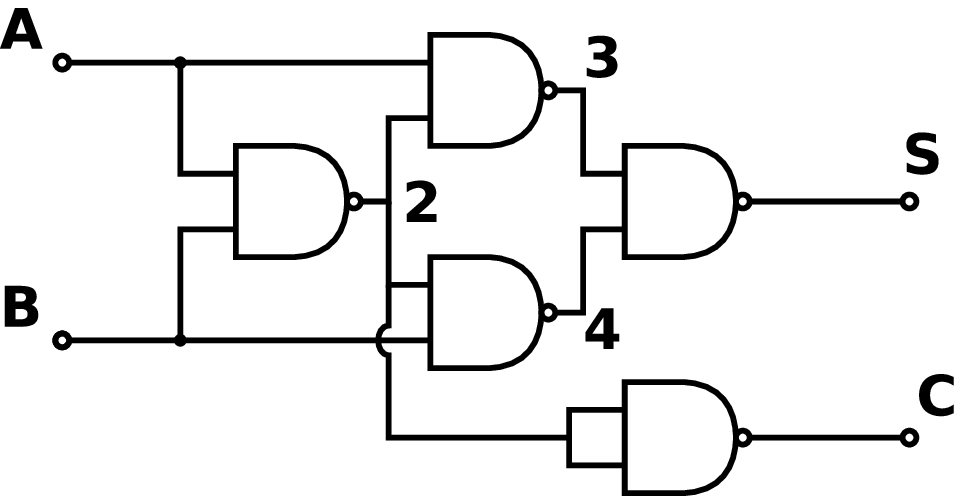}}
\subfigure[]{
\includegraphics[width=0.35\columnwidth]{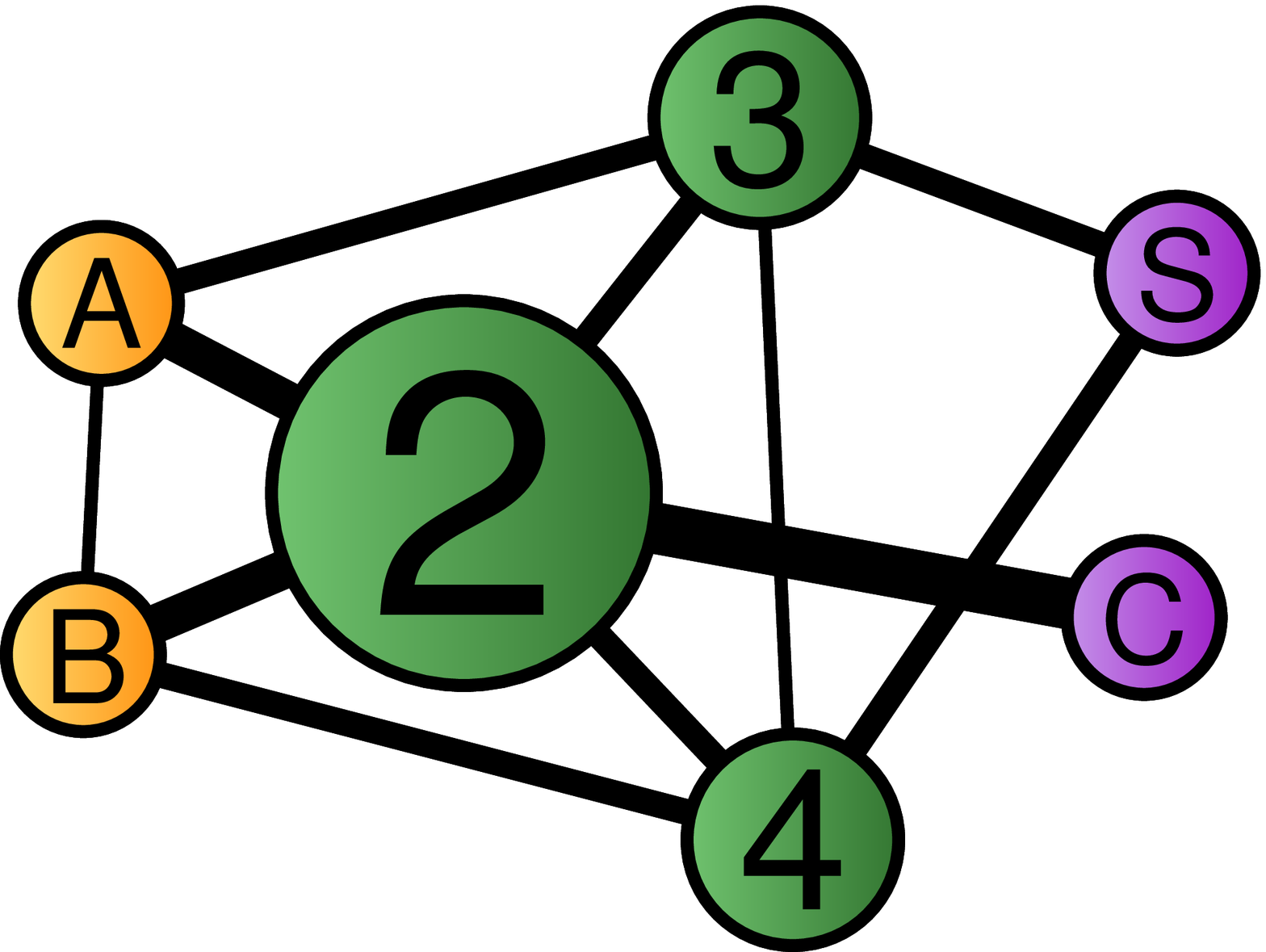}}\\
\caption{(Color online) Ground state embedding of the half adder circuit.
a) The half adder is implemented with a \XOR\ gate and an \AND\ gate.
b) The \XOR\ and \AND\ gates have been substituted by the 
corresponding all-\NAND\ circuits. c) The same circuit has been rewritten 
without the redundant gates and labeled wires. d) Here 
the circuit is mapped to a network of seven spins, each corresponding to the 
seven wires of the circuit.
The thickness of each link is proportional to the two-spin interaction strength, 
while the size of each node is proportional to the local field strength in 
the two-local reduction. 
The parameters used for the \NAND\ gate 
Hamiltonian given in eq.~\eqref{eq:hnand}
are $c_1=c_2=c_{12}=1$.}
\label{fig:halfadder}
\end{figure}

Let us turn to an illustration that shows how to use the Hamiltonian
in eq.~\eqref{eq:hnand} to construct more complex functions.
Naively, it may seem a separate spin must be included for each wire 
originating from a {\sf FANOUT}~operation~\cite{JDB08,Gu12,Crosson10}.
However, this is not the case;
instead the same spin may be used for the input to as many gates as desired.
As an example, in fig.~\ref{fig:halfadder}, an all-\NAND~half adder circuit is 
converted to a spin Hamiltonian using eq.~\eqref{eq:hnand}. 
We will return to this example at the end of the letter as an 
application of our symmetry considerations. 

An important consideration for this 
model is the input and output of the circuit.  
To extract data from this system, single spin projective measurements can be 
used. 
Inputs are set using an additional Hamiltonian 
\begin{equation}
H_{in}
=\frac{1}{2}\sum_{k}^{inputs} 
(\id + (-1)^{1-x_k}\sigma_k)
\label{eq:Hin}
\end{equation}
which forces the $k$-th bit to take the value $x_k\in\{0,1\}$.

 There are certain symmetries of Boolean
functions from which we can infer properties of the class of 
Hamiltonians that have the Boolean function embedded
in the ground state subspace. 

To limit the scope of our initial discussions, we will 
restrict our attention to Hamiltonians containing 
only two-spin interactions and to the set of the 
16 two-input, one-output gates. 

Each of the two-input, one-output gates is defined by its truth table:
$$
\begin{tabular}{cccc}
\hline
$x$ & $y$&\phantom{sp} &$z$ \\
\hline
0 & 0 && $b_1$  \\
0 & 1 && $b_2$  \\
1 & 0 && $b_3$  \\
1 & 1 && $b_4$  \\ 
\hline
\end{tabular}
$$
with $b_i\in\{0,1\}$.  There are 16 choices for the 
vector $b=[b_1,b_2,b_3,b_4].$
The corresponding Hamiltonian, $H_b$, must have ground state subspace
\begin{equation}
	\LL(H_b)=\{ \ket{00b_1},\ket{01b_2},\ket{10b_3},\ket{11b_4}\}
\end{equation}
Thus, there are 16 relevant ground state subspaces, each corresponding to 
one of the truth tables.

\begin{figure}[thb]
  \begin{center}
	  \includegraphics[width=\columnwidth]{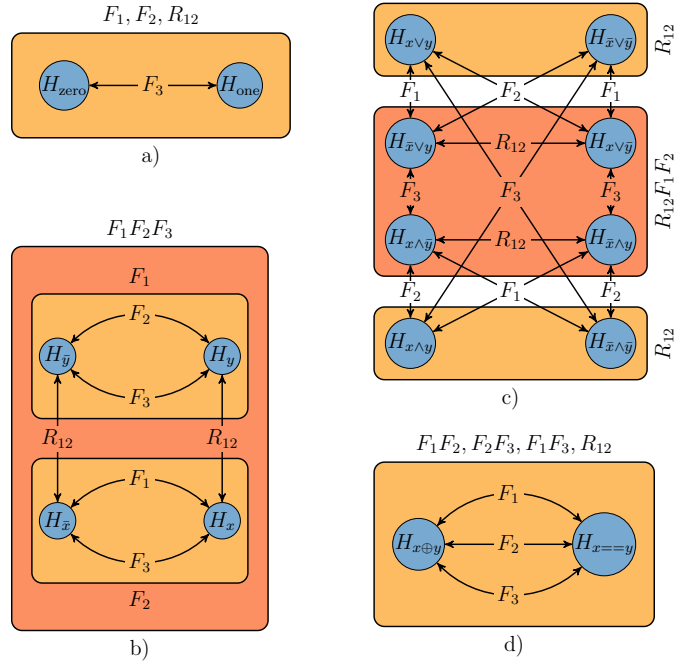}
  \end{center}
  \caption{(Color online)
  The action of $D_4\times Z_2$ on the 16 Hamiltonians corresponding to
  truth tables of two-input, one output functions. The Hamiltonians can 
  be converted to any other Hamiltonian in the same orbit by applying
  the spin-flip (negate) $F_i$ or input-swap (permute) $R_{12}$
  operations. The symmetry operations that leave the ground state subspaces of 
  each Hamiltonian invariant (the stabilizer subgroup)
  is written on the perimeter of each rectangular
  region. Orbits a), b), c), d) are explained separately in the text.
  Each of these orbits requires an additional spin for a Hamiltonian 
  embedding using only two-spin interactions: 
  orbit a) requires a single spin, b) two spins, c) three spins, and orbit d) requires four spins.} 
  \label{fig:classes}
\end{figure}
The symmetry operations on truth tables must treat the
output bit differently in order to remain in the space of the 16
truth tables. Thus, we consider (i) bit flips of any of the spins 
and (ii) swaps of the two inputs giving the following symmetries:
$\{e,F_1,F_2,F_3,R_{12}\}$. Here $e$ is the identity operation, $F_i$ is the 
spin-flip operation (negate), and  $R_{12}$ is the spin-swap operator (permute).
The action of the latter two operations on spins is defined via
\begin{eqnarray}
	F_i\circ\sigma_j&=& (1-2\delta_{ij})\sigma_j\label{eq:symF}\\
	R_{ij}\circ \sigma_k&=&\sigma_j\delta_{ki}+\sigma_i\delta_{kj}+\sigma_k(1-\delta_{ki}-\delta_{kj})\label{eq:symR}
\end{eqnarray}
The group $G$ can be presented as 
\begin{equation}
	G=\langle R_{12},F_1,F_3\rangle
	\label{symGrp}
\end{equation}
where $\langle \cdot\rangle$ indicates a set of generators. Defining
relations of the group are $R_{12}^2=F_1^2=F_3^2=e$, $R_{12}F_1=F_2R_{12}$,
$F_1F_3=F_3F_1$ and $R_{12}F_3=F_3R_{12}$. From these relations, or alternatively
from the cycle graph, the group is of order 16 and is isomorphic to 
$D_4\times Z_2$, where $D_4$ is the symmetry group of the square and
$Z_2$ is the cyclic group of order $2$.  

The action of $G$ on  the set of 16 truth tables
is depicted in fig.~\ref{fig:classes}. 
Four orbits are found under action of the group:
\begin{eqnarray*}
&&\{0,1\},\\ 
&&\{x,y,\bar{x},\bar{y}\},\\  
&&\{x\lor y,\bar{x}\lor y, x\lor\bar{y},\cdots,\bar{x}\land \bar{y}\}\\
&&\{x\oplus y, x == y\}. 
\end{eqnarray*}
These classes are depicted 
in fig.~\ref{fig:classes}a, \ref{fig:classes}b, \ref{fig:classes}c, \ref{fig:classes}d, 
respectively. These classes correspond to different NPN (negate-permute-negate) 
classes~\cite{Correia01,Chang99}.
Interestingly, each orbit requires a different
number of spins to implement when considering only two-spin interactions.
We examine each in turn. 

First, consider the constant functions with $b_i=c$ and $c\in\{0,1\}$. Since
these functions do not depend on $x$ nor $y$, there is no need to couple either
to the third spin. Hence, the Hamiltonian in eq.~\eqref{eq:Hin} can be used.
According to the group action depicted in fig.~\ref{fig:classes}a, 
given the Hamiltonian for $H_{zero}$ corresponding to $b_i=0$, the action of $F_3$ 
transforms $H_{zero}$ to $H_{one}$.

Second, for each of the functions, $b_i=x$, $b_i=y$, $b_i=\bar{x}$ and 
$b_i=\bar{y}$, the output bit only depends on one of the two inputs. The other
input is extraneous, so the gate only requires two spins to implement.
The truth tables can be embedded using variations of the
\COPY~gate previously introduced in~\cite{Gu12,JDB08,Crosson10}. The 
general $k$-\COPY\ gate forces $k$ bits to take the same value and the
corresponding diagonal operator
\begin{equation}
H_{k\textrm{-}\COPY}=
- \frac{1}{2}\sum_{i\neq j} \sigma_{i}\sigma_j
\end{equation}
acting on $k$-spins possesses a ground state subspace 
\begin{equation}
	\LL(H_{k\textrm{-}\COPY})=\Span\{\ket{0}^{\otimes k},\ket{1}^{\otimes k}\}.
\end{equation}
If we are concerned with constructing
a Hamiltonian using a physical set of spins, the spatial locality could play
an important role as coupling of distant spins may not be possible. In this
case, the $k$-\COPY~gate could be useful for spatially distributing 
intermediate results of the computation.  
The action of $F_1$ or $F_3$ transforms 
$H_{x}=H_{2\textrm{-}\COPY(x,z)}$ into the Hamiltonian $H_{\bar{x}}$, 
as shown in fig.~\ref{fig:classes}b.

The third class of functions to be considered is $x\lor y$, $x\land y$ and all 
possible negations of the two inputs. Our general formula for 
$\bar{x}\lor\bar{y}$ is given in eq.~\eqref{eq:hnand} and using the symmetry operations 
from group $G$, see fig.~\ref{fig:classes}c, all other gates in this orbit can be 
derived using three spins with two-spin interaction terms (see the appendix for additional formulations). 

The last orbit of functions, \XOR\ ($x\oplus y$) and its logical negation \EQUIV\ ($x==y$), 
cannot be embedded in the ground state subspace of a three spin system using only 
two-spin interactions; it requires a fourth ancilla spin to implement using only pairwise interactions. If restricted to three spins, the gate 
$\XOR$ ($\oplus$) requires a three-spin interaction.
\begin{equation}
	H_{x\oplus y}(\sigma_1,\sigma_2,\sigma_3)= -\sigma_1\sigma_2\sigma_3
\end{equation}
The inability to create this operator acting on  three spins with 
two-spin interactions can be demonstrated algebraically or graphically using
Karnaugh maps~\cite{JDB08,Rosenbaum}. 
For \XOR, the stabilizer subgroup is generated by 
$F_iF_j$ and $R_{12}$, see fig.~\ref{fig:classes}d.
When considering the ancilla spin, 
$\sigma_4$, there is an additional $F_4$ symmetry that leaves the
truth table unchanged.

Beginning with the swap-symmetric operators $M_z=\sum_i\sigma_i$
and $M_{zz}=\sum_{i<j}\sigma_i\sigma_j$, we write
the most general swap-symmetric Hamiltonian over four spins restricted to two-spin 
interactions as
\begin{equation}
	H_R=r_{z}M_z+r_{zz}M_{zz}+\sigma_4(r_{4}+r_{z4}M_z).
	\label{eq:rotsymH}
\end{equation}

Suppose that the coefficient vector $R=[r_z,r_{zz},r_4,r_{z4}]$ gives a valid \XOR\
Hamiltonian. Then we can act with $F_4$ to get a second Hamiltonian that also
preserves the ground state subspace with coefficients $R'=[r_z,r_{zz},-r_4,-r_{z4}]$.  In 
references~\cite{JDB08} and~\cite{Gu12}, this $F_4$ symmetry connects the
decompositions 
given as $R=[1,-1,-2,2]$ and $R=[1,-1,2,-2]$ in the respective papers. 
Furthermore, since the ground state subspace is symmetric 
with respect to $F_iF_j$, there are an additional six Hamiltonians with 
logically equivalent ground state subspaces. For example, beginning with $H_{x\oplus y}$
corresponding to $R=[1,-1,-2,2]$ and using symmetry operation $F_1F_2$ results in
\begin{eqnarray}
	F_1F_2\circ H_{x\oplus y}&=& 2\sigma_4(-\sigma_1-\sigma_2+\sigma_3)\\
			   &&+\sigma_1+\sigma_2-\sigma_3-2\sigma_4\nonumber\\
	&&+(\sigma_1\sigma_2-\sigma_2\sigma_3-\sigma_1\sigma_3)\nonumber
\end{eqnarray}
with the same ground state subspace.  Note that this Hamiltonian is not of the same form 
of eq.~\eqref{eq:rotsymH} like those given in~\cite{JDB08,Gu12}.

To extend the \XOR~Hamiltonians previously listed to a parameterized family of
Hamiltonians, we rearrange eq.~\eqref{eq:rotsymH} with $R=[1,-1,-2,2]$ as
\begin{eqnarray}
	\label{eq:xorA}
	H_{x\oplus y}&=& -(\sigma_1+\sigma_2)(\id-\sigma_4)\\
	         && -2\sigma_4+(\sigma_1\sigma_2+\sigma_1\sigma_4+\sigma_2\sigma_4)\nonumber\\
		 && -\sigma_3+\sigma_1\sigma_3+\sigma_2\sigma_3+2\sigma_3\sigma_4\nonumber
	\label{eq:xor2}
\end{eqnarray}
Comparing with eq.~\eqref{eq:hnand} and using fig.~\ref{fig:classes}c, we can simplify 
this equation using 
$H_{\bar{x}\land \bar{y}}(\sigma_1,\sigma_2,\sigma_4)=F_1F_2F_4\circ H_{{\bar{x}\lor\bar{y}}}(\sigma_1,\sigma_2,\sigma_4)$
evaluated at $c_1=c_2=c_{12}=1$.  
Generalizing to other values of $c_1,c_2,$ and $c_{12}$, we arrive at the following 
three-parameter family that preserves the ground state subspace of \XOR
\begin{eqnarray}
	H_{x\oplus y}&=& H_{\bar{x}\land \bar{y}}(\sigma_1,\sigma_2,\sigma_4)-\sigma_3\\
		 && +\sigma_1\sigma_3+\sigma_2\sigma_3+2\sigma_3\sigma_4\nonumber
	\label{eq:xor3}
\end{eqnarray}

\begin{table}[!t]
\ 
\vskip 5pt
 \begin{tabular}{l l}
 \toprule
 $z=f(x,y)$ & $H_{f(x,y)}(\sigma_1,\sigma_2,\sigma_3,\sigma_4)$\\
\hline
\multicolumn{2}{l}{Constant functions}\\
$z=0$   &$H_{zero}=(\id-\sigma_3)$\\
\multicolumn{2}{l}{Copy-type functions}\\
$z=x$ &$H_{x}=(\id-\sigma_1\sigma_3)$\\
\multicolumn{2}{l}{\AND, \OR, \dots, \NAND, \NOR~functions}\\
$z=\bar{x}\lor\bar{y}$&$\begin{array}{ll}H_{\bar{x}\lor \bar{y}}=&(c_1\sigma_1+c_2\sigma_2)(\id+\sigma_3)\\
	&+(c_1+c_2)\sigma_3+c_{12}\sum_{i<j}^3\sigma_i\sigma_j\end{array}$\\
\multicolumn{2}{l}{\XOR~and \EQUIV~functions}\\
$z=x\oplus y$ & $
\begin{array}{ll}H_{x\oplus y}=&H_{\bar{x}\land \bar{y}}(\sigma_1,\sigma_2,\sigma_4)-\sigma_3\\
	&+\sigma_1\sigma_3+\sigma_2\sigma_3+2\sigma_3\sigma_4\end{array}$\\
\bottomrule
\end{tabular}
\caption{Summary of representative Hamiltonians from each orbit under the
action of the symmetry group. Spin one and two correspond to the two inputs while
spin three corresponds to the output.  The fourth spin is an ancilla
spin needed only for the implementation of $\XOR$ and $\EQUIV$.
In the \AND, \OR, \dots, \NAND, \NOR~family, the sign of the coefficients 
determines which gate on this NPN orbit one obtains, as detailed in the appendix.
We have only shown four Hamiltonians and the remaining
12 Hamiltonians as well as additional Hamiltonians with different excited states
are related via the action of the group $D_4\times Z_2$ as depicted in 
fig.~\ref{fig:classes}.}
\label{tbl:3to2}
\end{table}
By examining the excited state structure of eq.~\eqref{eq:xorA}, 
we find that in the parameterization of $H_{\bar{x}\land\bar{y}}$ the coefficients, $c_1,c_2,c_{12}$, must
be greater than $1/2$ instead of strictly positive.

Our work has 
direct relevance to recent experimental realizations of 
adiabatic quantum computing in superconducting 
qubits~\cite{2010PhRvB..82b4511H,Johnson11} and ion 
traps~\cite{Britton12} where controllable couplings between spins can be used
to embed problems into the target Hamiltonian of the evolution. Since both of
these experimental systems are limited to two-spin interactions, our
decomposition for \XOR\ provides an effective three-spin interaction
which is experimentally realizable.

In table~\ref{tbl:3to2}, we summarize our results for Hamiltonian embeddings 
of two-input, one-output Boolean functions.
While we have restricted attention to diagonal Hamiltonians,
future work could consider transformations where the ground state is preserved
but the Hamiltonian obtains off-diagonal elements. 

\begin{figure}[!t]
  \begin{center}
    \includegraphics[width=0.2\textwidth]{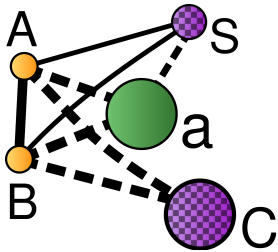}
  \end{center}
  \caption{(Color online) The half adder spin Hamiltonian that arises from the  four spin decomposition 
  of the \XOR\ Hamiltonian which simplifies the construction from fig.~\ref{fig:halfadder}d. Dashed links represent
  negative interactions and checkerboard shading indicates a negative local field. The size 
  of the nodes and the thickness of the edges are proportional to the fields 
  and interaction strength (on spins $A$ and $B$ there is no local field). The 
  parameters used for the \AND~gate and \XOR~gate are $c_1=c_2=c_{12}=1$.} 
    \label{fig:halfadder-xor}
\end{figure}
Now we return to the half adder example from fig.~\ref{fig:halfadder}. With our constructions, 
we can directly implement it using the \XOR\ and \AND\ gates,
\begin{equation}
	H_{HA}=H_{x\oplus y} (\sigma_A,\sigma_B,\sigma_a,\sigma_S)+
	H_{x\land y}(\sigma_A,\sigma_B,\sigma_C).
\end{equation}
Here $\sigma_A$ and $\sigma_B$ correspond to the inputs to be summed,
$\sigma_a$ corresponds to the \XOR\ ancilla bit, and $\sigma_S$ and $\sigma_C$ 
correspond to the sum and carry bits. As depicted in fig.~\ref{fig:halfadder-xor}, 
the new spin Hamiltonian uses two less ancilla spins than our earlier construction
and now has six free parameters. Additional degrees of freedom arise from the $D_4$
stabilizer subgroup of the \XOR\ Hamiltonian and the $Z_2$ stabilizer subgroup of the
\AND\ Hamiltonian.

\begin{figure*}[!htb]
\centering
\subfigure[]{
\includegraphics[width=0.48\textwidth]{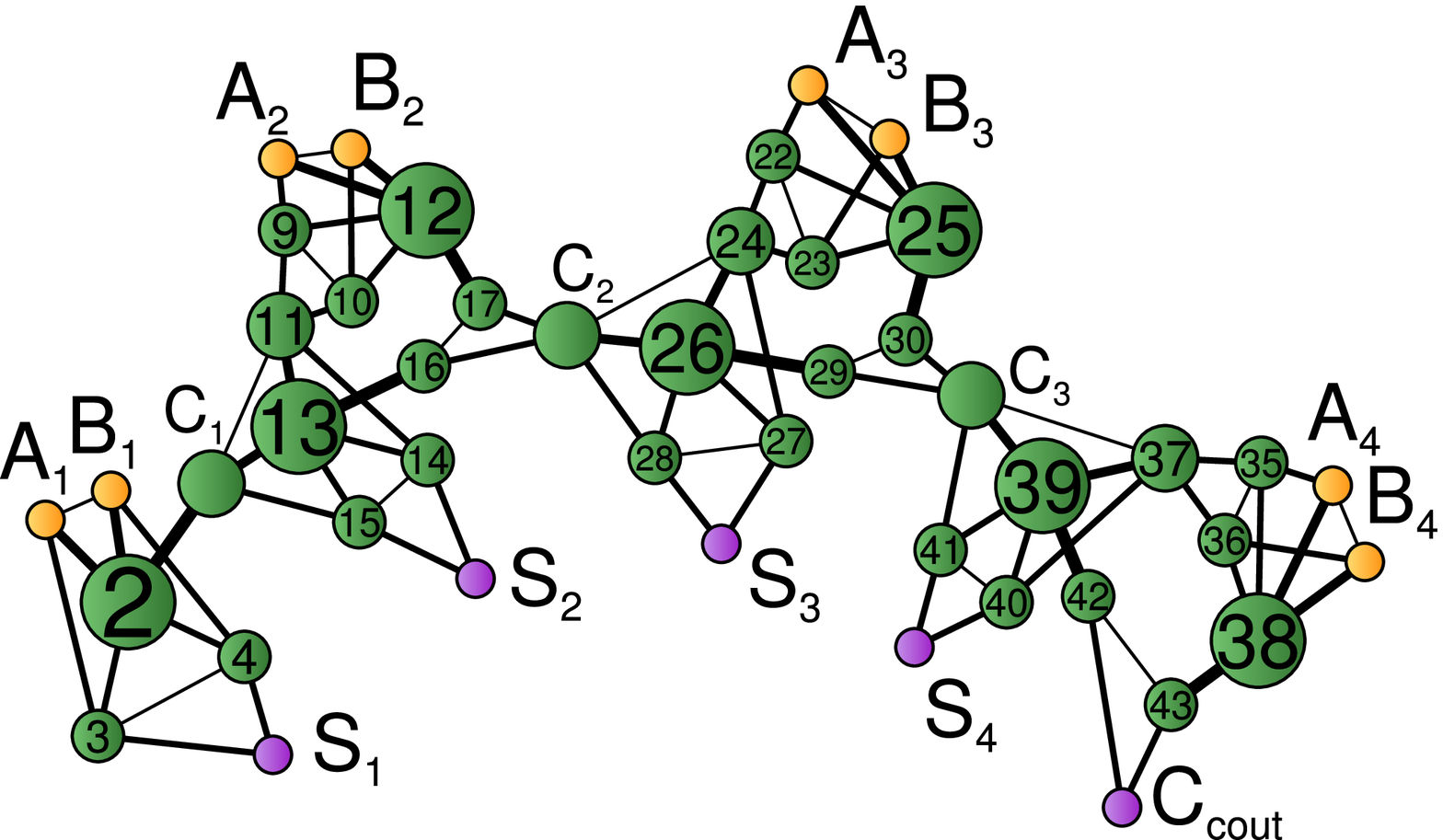}
\label{fig:ripple-nand}
}
\subfigure[]{
\includegraphics[width=0.48\textwidth]{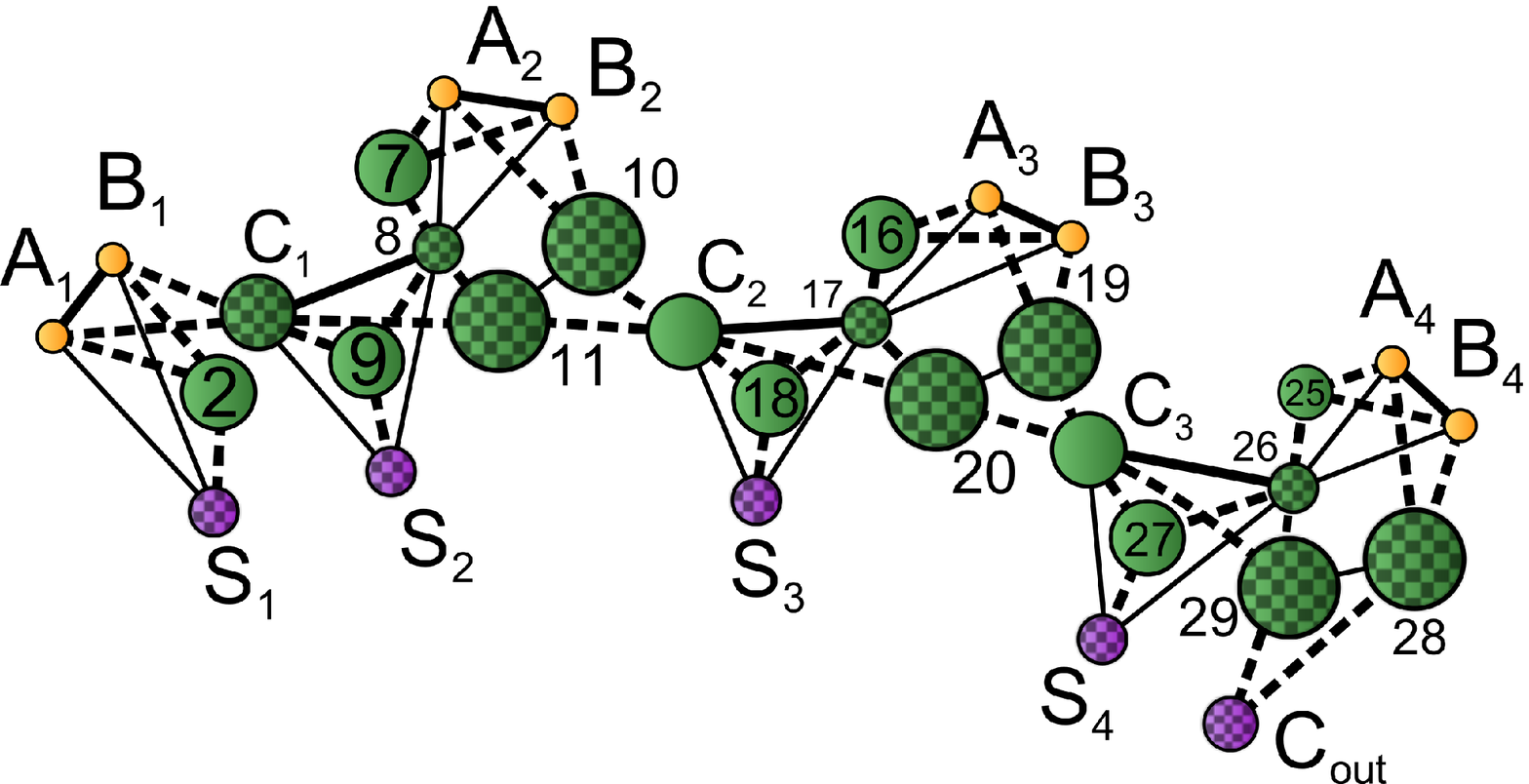}
\label{fig:ripple-xor}
}
\caption{(Color online) Ripple carry adder. The figure shows the network of spins corresponding
to a ripple carry adder with four bits.
The ripple carry adder is composed by one half adder and three full adders; in
yellow it shows the input spins from the four bits binary numbers 
$A=\sum_{i=1}^4A_i2^i$ and $B=\sum_{i=1}^4B_i2^i$;
while the sum spins, $S_i$ are drawn in purple. Carry bits are labeled as $C_i$.
The direction of the sum is from left to right.
Fig.~$a$) shows a ripple carry constructed with only \NAND\ gates and
parameters $c_1=c_2=c_{12}=1$, while $b$) 
shows the same adders built with
\XOR, \AND\ and \OR\ gates.
} 
\label{fig:ripplecarry}
\end{figure*}

The symmetry group of $H_{HA}$ can be inferred from the symmetries of the
component Hamiltonians using a direct product structure. 
For a general circuit Hamiltonian composed of gate Hamiltonians acting
on subsets of spins, $H=\sum H_i$, 
the stabilizer subgroup is the direct product of the stabilizers for each of the 
Hamiltonians in the sum. The direct product group action is defined as 
$(g_1,g_2,\cdots g_N)\circ H=\sum g_i\circ H_i$. If $g$ is in the
intersection of all stabilizer groups (the diagonal subgroup),
then $g\circ H$ will have the same ground state subspace as $H$. 

Additional symmetries arise after partitioning the bits into output and 
ancilla bits.  We can expect the symmetries of the Boolean function being embedded
to be possessed by the resulting Hamiltonian. However, the symmetry group
composed of the gate-local symmetries preserves the full ground state 
subspace including the values of the ancilla bits.  The symmetries
of the Boolean function before being decomposed into logic gates will 
arise as global symmetries that cannot be obtained from the gate-local 
symmetries of the individual gates. For instance, if $\sigma_a$ 
corresponds to an ancilla spin, then inverting this bit in each circuit
component leaves the ground state subspace invariant. That is, 
$H$ and $(F_a,F_a,\cdots,F_a)\circ H$  embed the same Boolean function. 

As a further illustration of the distinction between global and gate-local 
symmetries, consider the full adder 
corresponding to a Boolean function which adds binary summands $A$, $B$, and 
carry-in bit $C_{in}$. The permutation of the input bits 
and the carry-in bit is a symmetry of the full adder
Boolean function. However, such a permutation is not a gate-local symmetry 
of the sub-Hamiltonians used in the circuit embedding, see the appendix for details.
This is because the values of the ancilla spin within the ground state subspace 
is not preserved under this permutation. Thus, the local symmetries do 
not determine all possible symmetries when some bits are considered as ancillas. 

As a final example of ground state spin logic, fig.~\ref{fig:ripplecarry} 
shows the spin 
Hamiltonian of the ripple carry adder for four-bit binary numbers.
The figure shows the network for both an implementation with only \NAND\ gates
in fig.~\ref{fig:ripple-nand}
and an implementation with \XOR, \AND, and \OR\ gates in fig.~\ref{fig:ripple-xor}. 
The second construction 
allows a decrease in the number of ancilla spins and provides $51$
free parameters. Additionally, as shown in the appendix, the symmetry group of the second
implementation has at least $2^{31}$ elements. Another salient feature is that the 
average degree of the spins changes from $3.85$ in the all-\NAND\ case to $4.22$ in
the second implementation.
Explicitly listing the free parameters and the symmetries that preserve the 
ground state subspace is an illustration of how our approach gives 
experimentalists and theorists systematic methods
to find additional degrees of freedom.

An important step towards large scale experimental realizations of the techniques presented in this paper will be the adiabatic
implementation and characterization of the elementary logic gates.  In the case of  
\XOR, this Hamiltonian will allow one to realize an effective
three-spin interaction by using only two-spin interactions and introducing an
ancilla spin.  Such an interesting example is
in line with current experimental capabilities \cite{Britton12,Johnson11,  2010PhRvB..82b4511H}.

\acknowledgements
The authors would like to thank V. Bergholm and Z. Zimboras for 
helpful discussions 
and M. Allegra and J. Roland for carefully reading the manuscript.
JDW acknowledges support from NSF (No.~1017244)
and thanks the Visitor Program at the Max-Planck Institute for 
the Physics of Complex Systems, Dresden where parts of 
this work were completed.

\input{appendix}

\end{document}

%% file: appendix.tex
\section{Appendix}

\subsection{Hamiltonians embedding full adders}
We provide the characterization of the full adder~\cite{Hayes} necessary to
construct the ripple carry adder shown in the main text.
This affords us an opportunity to 
explore the network properties of the adders circuit family with well known
constructions and optimized solutions~\cite{Vespignani,Wegener}.

\begin{figure}[thb]
\centering
\subfigure[\ Full adder circuit]{
\label{fig:full-xor-circuit}
\includegraphics[width=0.48\columnwidth]{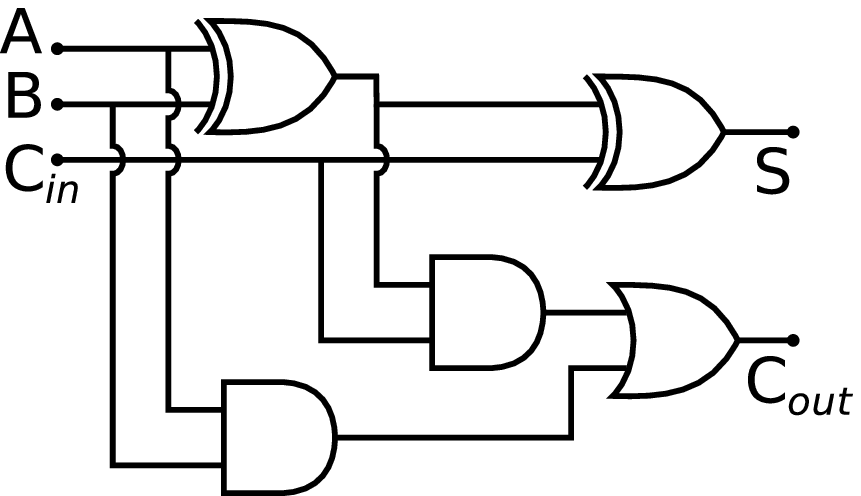}}
\subfigure[\ Full adder circuit with \NAND s]{
\includegraphics[width=0.48\columnwidth]{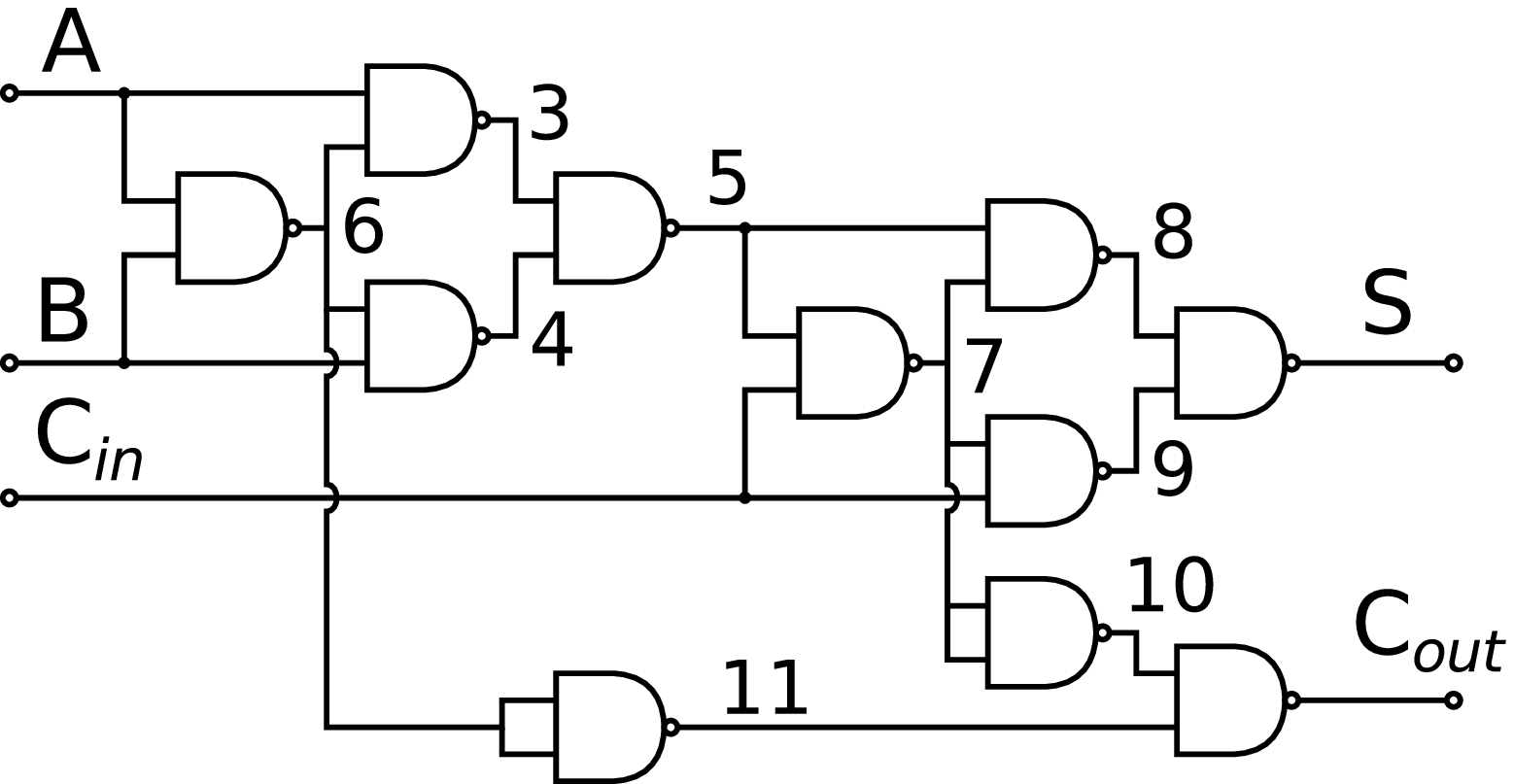}}
\subfigure[\ Full adder network]{
\includegraphics[height=0.48\columnwidth]{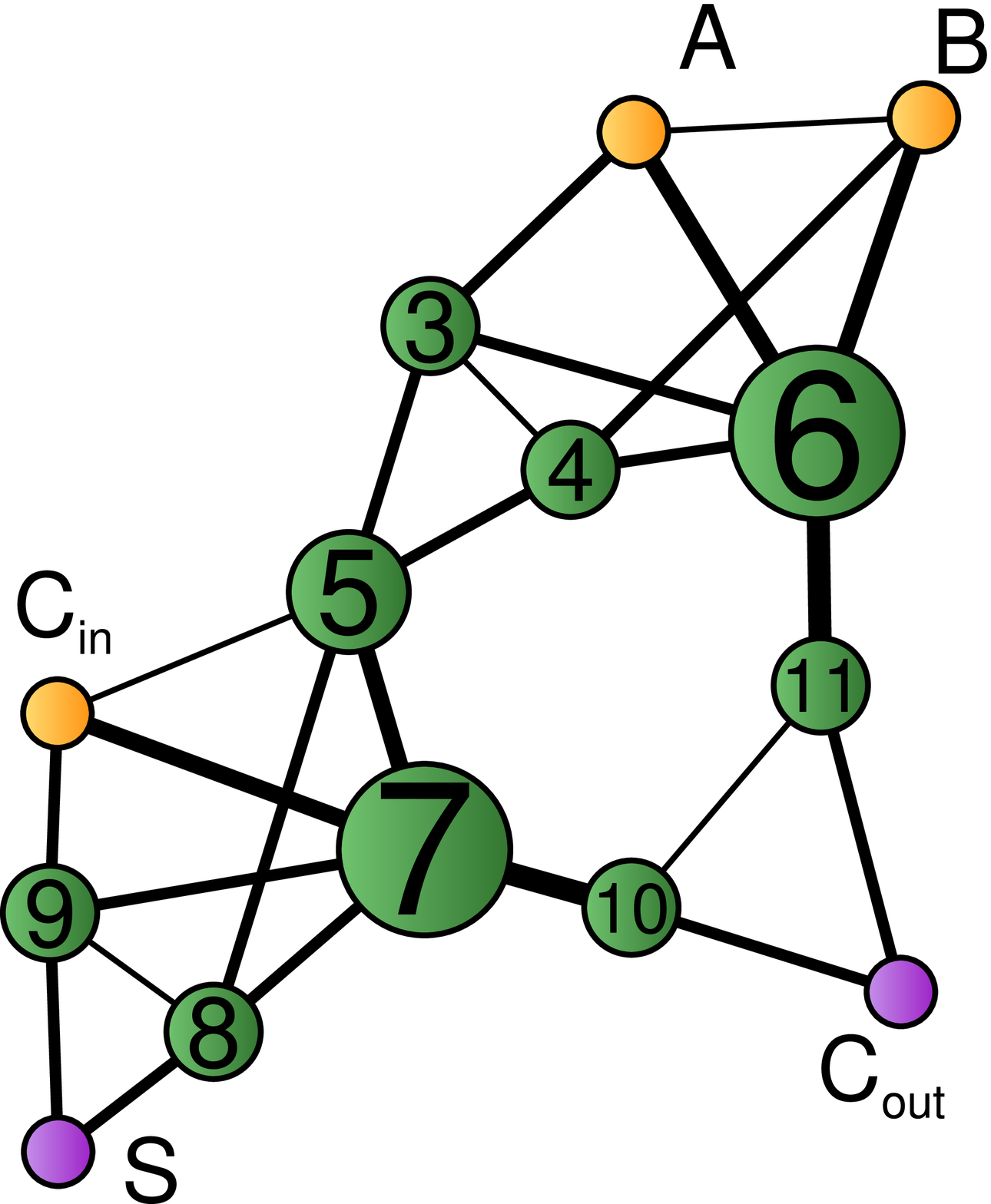}}
\subfigure[\ Centrality of nodes]{
\includegraphics[height=0.48\columnwidth]{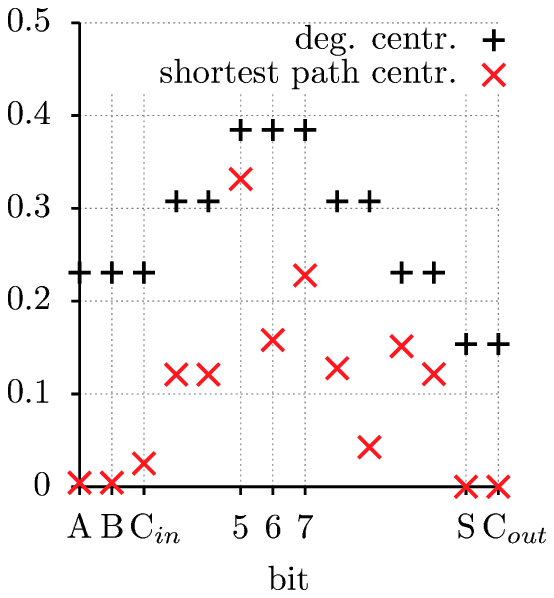}
\label{fig:fulladder_centrality}}
\caption{(Color online) a) Full adder circuit.
b) Transcription with only \NAND\ gates.
c) The spin network representing the all-\NAND\ circuit of the full adder.
The input spins $A$, $B$ and the lower carry-in spin $C$ are depicted in yellow
while the output spins of sum ($S$) and carry-out ($C_{out}$) are colored in purple.
The green nodes represent ancilla spins.
d) Degree centrality and shortest path centrality of network nodes.
Notice that the input and output spins (denoted $A=0$, $B=1$, $C_{in}=2$ and 
$S=12$, $C_\textrm{out}=13$ respectively)
are the least central while the most central nodes are the ancilla spins
five, six and seven.}
\label{fig:fulladder}
\end{figure}

In order to sum arbitrarily large binary numbers, the half adder circuit needs to
implement the bit carrying operation. 
The full adder circuit introduces this operation with a
third input bit, accounting for the lower level carry bit $C_{in}$.
Fig.~\ref{fig:fulladder}\ shows the network associated to this circuit,
where the inputs bits $A,B$ and $C_{in}$ are in yellow and the output bits $S$ and
$C_{out}$ are in purple.
The network corresponds to a circuit with only \NAND\ gates. The two-spin
interactions and the local fields are then all positive valued.
From eq.~\eqref{eq:hnand}, we have three free parameters for
each \NAND\ gate, giving $3\cdot5=15$ free parameters.
Each \NAND\ Hamiltonian is also symmetric under the action of the symmetry group
$\{e,R_{12}\}$, giving a symmetry group for the whole Hamiltonian of at least
$2^5$ elements.
The Hamiltonian uses nine ancilla spins to build the truth table of the
full adder, resulting in nine new symmetries, labelled in the main text as
$\{F_a: a\textrm{ labels an ancilla spin}\}$. The action of the latter changes
the ground state subspace but the resulting system still describes the original problem.

\begin{figure}[!htb]
  \begin{center}
    \includegraphics[width=0.5\columnwidth]{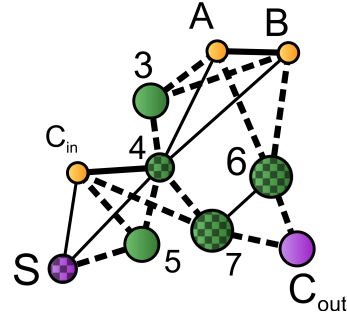}
  \end{center}
  \caption{(Color online) Full adder spin network for the circuit in
  fig.~\ref{fig:full-xor-circuit}.
  This construction reduces the number of ancilla spins from nine in the case in
  fig.~\ref{fig:fulladder_centrality} to five.
  In this case, the interactions are not all of the same sign.}
  \label{fig:full-xor}
\end{figure}

The number of ancilla spins can decrease using the standard \XOR, \AND\ and \OR\ 
gates of fig.~\ref{fig:full-xor-circuit}.
Fig.~\ref{fig:full-xor} shows the spin network associated to this circuit.
This Hamiltonian presents at least $2^9$ symmetries arising from the single-gate
symmetries.

To enhance comprehension of the spin Hamiltonians,
we compute some well known complex networks measures~\cite{Vespignani}.
The node centrality of the resulting network, in 
fig.~\ref{fig:fulladder_centrality}, suggests that the input and
outputs spins are the least central both for local (degree
centrality) and global centrality measures (shortest path centrality).
Here the degree centrality $D_k$ of node $k$ is defined as:
\begin{equation}
D_k=\frac{d_k}{N-1},
\label{eq:dc}
\end{equation}
where $d_k$ is the degree of node $k$ and $N$ is the number of nodes of the
network. The shortest path centrality is defined as:
\begin{equation}
  SP_k=\sum_{i,j}\frac{SP_{ikj}}{SP_{ij}},
  \label{eq:spc}
\end{equation}
where $SP_{ij}$ represents the number
of shortest path between nodes $i$ and $j$ and $SP_{ikj}$ is the number of
those paths passing through node $k$.  
Nodes with higher centrality can be thought as the network bottleneck between input
and output spins.
\subsection{Additional calculations for the ripple carry adder}
The sum of two binary $n$-bit numbers $x$ and $y$, can be carried out 
by concatenating $n$ full adder circuits yielding the ripple carry
adder.
This circuit implements a cascade: the carry-out bit of each full 
adder will be used as the carry-in bit for the next one.
The first full adder has logical zero as the carry-in bit. Alternatively,
it can be completely replaced by the half adder circuit.
Fig.~\ref{fig:ripplecarry} of the main text, shows the network associated to a four bit
ripple carry adder used to sum two four-bit binary numbers.
In fig.~\ref{fig:ripple-nand} the Hamiltonian is built from \NAND\ gates, providing
all non-negative local fields and interactions and resulting in 46 spins and 86
links.
The starting circuit contains 38 \NAND s, each of them are associated to a
Hamiltonian, see eq.~\eqref{eq:hnand}, which depends on three free
parameters. Thus, the Hamiltonian of the whole circuit has $3\cdot38=114$
free parameters.
As in fig.~\ref{fig:classes}, the stabilizer subgroup of the \NAND\ gate contains only
two elements:
\begin{equation}
  \textrm{stab}(\NAND)=\langle R_{12}\rangle \simeq Z_2
\end{equation}
and generates a group of symmetries for the full Hamiltonian of at least $2^{38}$ elements.
Fig.~\ref{fig:ripple-xor} shows the same circuit built using seven \XOR, seven
\AND\ and three \OR\ gates. 
This implementation yields a network with only 32 bits and 65 links. 
The gate Hamiltonians have three free parameters each 
for a total of $3\cdot17=51$ free parameters.
In this case, the stabilizer subgroup of both \OR\ and \AND\ is also generated by $R_{12}$, while for the \XOR\ gate we have:
\begin{equation}
  \textrm{stab}(\XOR)= 
  \langle F_1F_2,F_1F_3,R_{12}\rangle\simeq D_4
\end{equation}
with eight elements. Thus, total symmetry group of the Hamiltonian contains at least
$2^{10}\cdot8^7=2^{31}$ elements.

\begin{figure*}[!htb]
\centering
\subfigure[\ Ripple carry network with only \NAND\ gates]{
\includegraphics[width=0.9\textwidth]{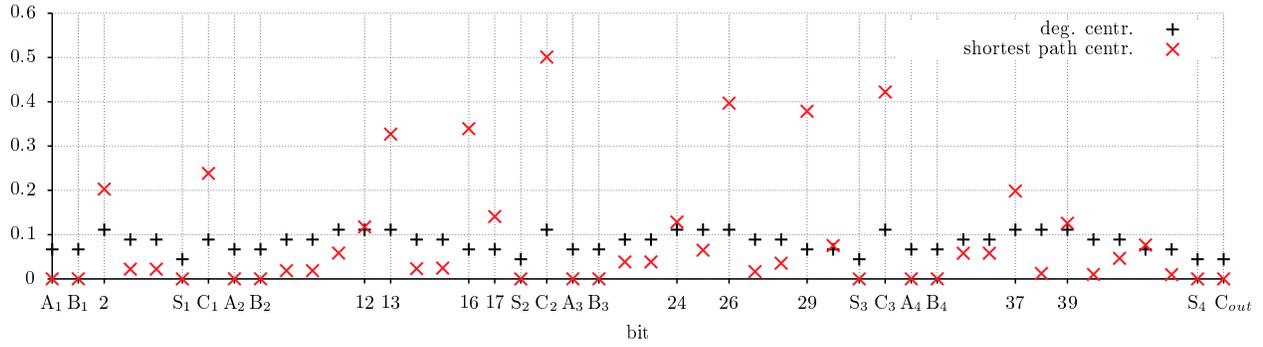}
}
\subfigure[\ Ripple carry network using standard gates]{
\includegraphics[width=0.9\textwidth]{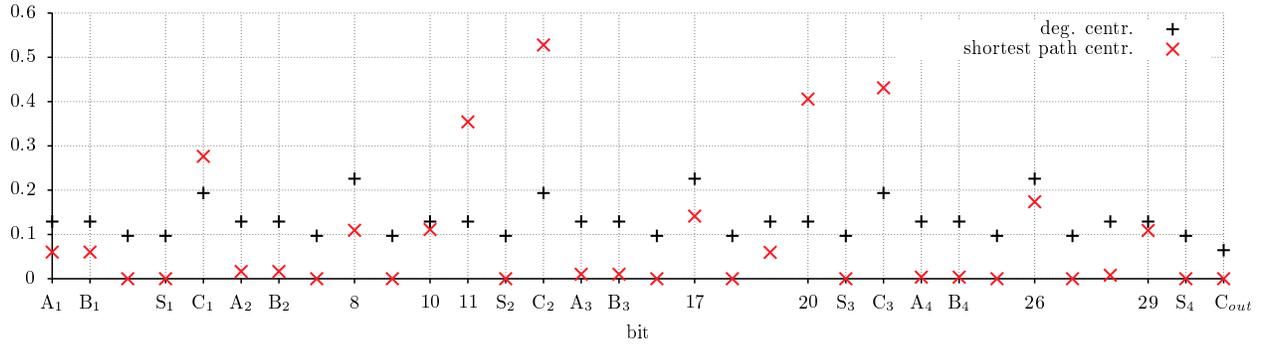}
}
\caption{(Color online) Centrality measures for the ripple carry adders of 
fig.~\ref{fig:ripplecarry}.
Graphs a) and b) correspond to networks in
fig.~\ref{fig:ripple-nand} and \ref{fig:ripple-xor} respectively.  
In the first implementation, the number of spins used is 46 with an average
degree centrality of 0.083, while in the latter the same problem is embedded
with a lower number of spins, 32, but with a higher average degree centrality of 0.131.
} 
\label{fig:ripple-centr}
\end{figure*}

Fig.~\ref{fig:ripple-centr}\ shows the centrality of each spin of the networks
in fig.~\ref{fig:ripplecarry}.
We note that, in the second implementation of the ripple carry adder, 
the resulting network is slightly more connected. The average degree centrality
($\langle D_i\rangle=0.131$) in this case is higher
than in the implementation with only \NAND s ($\langle
D_i\rangle=0.083$). 
The variance is also higher in the second example,
$\textrm{var}(D_i)=0.041$, than in the first, $\textrm{var}(D_i)=0.021$.
In both implementations, the most important spins as identified by the global measure
of shortest path centrality are the spins on the backbone of the
circuit. In particular, the carry bits, $C_i$, have high centrality as they connect
subnetworks which would otherwise be disconnected.

\subsection{General formulas for orbit of \NAND\ under $D_4\times Z_2$}  

From the main text, the Hamiltonian for \NAND\ is 
$$ 
H = (c_1 \sigma_1 + c_2 \sigma_2)(1+\sigma_3) + (c_1 + c_2)\sigma_3 + c_{12}M_{zz}
$$

This can be written as 
\begin{eqnarray}
H &=&  c_{12}(\sigma_1\sigma_2 + \sigma_1\sigma_3 + \sigma_2\sigma_3) \\
&&+ \sigma_3(c_1(1+\sigma_2(1+\sigma_3))) \nonumber\\
&&+ \sigma_3(c_2(1+\sigma_2(1+\sigma_3))) \nonumber
\end{eqnarray}
The energy shift to ensure that the ground state is also the null space is 
$$ 
c_1 + c_2 + c_{12}.
$$ 
We see that the symmetry in variables $\sigma_1$ and $\sigma_2$ breaks for
$c_1\neq c_2$, yet the ground state subspace remains invariant.  As mentioned in
the main text, this degree of freedom could be desirable as it reduces the
constraints placed on an experimental realization.

We can identify \NAND\ as a point on the orbit of the NPN class by considering
three indicator variables, $x,y,z\in \{0, 1\}$.  We write
\begin{eqnarray*}
H &=&  ((-1)^xc_1 \sigma_1 + c_2 (-1)^y\sigma_2)(1+(-1)^z\sigma_3)  \\\nonumber
&&+ (-1)^z(c_1 + c_2)\sigma_3 + c_{12}((-1)^{x+y}\sigma_1\sigma_2\\\nonumber
&&+  (-1)^{x+z}\sigma_1\sigma_3 + (-1)^{y+z}\sigma_2\sigma_3) 
\end{eqnarray*}

The Hamiltonian for \NAND\ is recovered by setting $x=y=z=0$.  The rest of the
orbit is picked out by assigning other values to $x,y,z$.  The ground state
energy as a function of $x,y,z$ is given as 

\begin{eqnarray*}
E_{gs} &=&  c_{12}( (-1)^{x+y} - (-1)^{x+z} - (-1)^{y+z})   \\\nonumber
&&- (c_1 + c_2)(-1)^z +\\
&&+ (1-(-1)^z)((-1)^xc_1 + (-1)^y c_2) \nonumber
\end{eqnarray*}

A meaningful experimental demonstration showing the capabilities to realize
every gate in the orbit could be performed by realizing each of the eight Hamiltonians $x,y,z\in \{0, 1\}$.
Apart from characterizing the degenerate ground space of each of the eight gates, the experiments can also be modified slightly to correspond to instances of adiabatic search algorithm as follows.  
The output of each gate can be set to either logical-zero or logical-one, see eq.~\eqref{eq:Hin}.  Successful adiabatic annealing would then return the associated inputs to the circuit.